\documentclass[iop,numberedappendix,twocolappendix,apj]{emulateapj}

\usepackage{natbib}
\usepackage{amsmath}
\usepackage{graphicx}
\usepackage{dcolumn}
\usepackage{bm}
\shorttitle{Analytical solutions of a fractional diffusion-advection equation}
\shortauthors{Litvinenko \& Effenberger}

\begin{document}

\title{Analytical solutions of a fractional diffusion-advection
  equation for solar cosmic-ray transport}

\author{Yuri E.~Litvinenko\altaffilmark{1} and
  Frederic Effenberger\altaffilmark{1}}
  \altaffiltext{1}{Department of
  Mathematics, University of Waikato, P.B. 3105, Hamilton, New
  Zealand. Email: yuril@waikato.ac.nz}

\begin{abstract}
  Motivated by recent applications of superdiffusive transport models
  to shock-accelerated particle distributions in the heliosphere, we
  solve analytically a one-dimensional fractional diffusion-advection
  equation for the particle density. We derive an exact Fourier
  transform solution, simplify it in a weak diffusion approximation,
  and compare the new solution with previously available analytical
  results and with a semi-numerical solution based on a Fourier series
  expansion. We apply the results to the problem of describing the
  transport of energetic particles, accelerated at a traveling
  heliospheric shock.  Our analysis shows that significant errors may
  result from assuming an infinite initial distance between the shock
  and the observer.  We argue that the shock travel time should be a
  parameter of a realistic superdiffusive transport model.
\end{abstract}


\section{Introduction}
When energetic particles, accelerated in the solar corona or the solar
wind, propagate in a turbulent heliospheric medium, the particle
transport is often diffusive \citep{Parker-1965}. 
The diffusion approximation is a standard tool in the description 
of evolving cosmic-ray distributions \citep[e.g.][and references 
therein]{Schlickeiser-Shalchi-2008,Artmann-etal-2011}.

A generalization of the diffusion equation is obtained by replacing
the usual derivatives by fractional derivatives. Formally, standard
diffusion, characterized by a linear growth of the variance of a
particle displacement, separates the processes of faster
superdiffusion and slower subdiffusion
\citep[e.g.,][]{Saichev-Zaslavsky-1997}. These processes are governed
by partial differential equations with fractional operators
\citep{Samko-etal-1993}. Physically, fractional differential equations
conveniently describe stochastic transport when an effective mean free
path is comparable with a macroscopic length scale. The corresponding
diffusive process is non-local: it is described by an integral
equation that can be rewritten in terms of fractional derivatives
\citep[for a clear discussion, see][]{Chukbar-1995}.

In sharp contrast to classical diffusion, solutions to fractional
diffusion equations typically are not Gaussian but rather have
power-law tails.  The key question in concrete applications is whether
the approach only provides a phenomenological parametrization for the
data or the postulated anomalous diffusion is based on physical
processes that can be described by a fractional evolution equation
\citep[see, e.g.,][for numerous potential
applications]{Metzler-Klafter-2000,Perrone-etal-2013}.

Observed solar cosmic-ray particle distributions often appear to
exhibit power-law tails, suggesting an interpretation in terms of
superdiffusion. Distributions of electrons and protons,
accelerated both in the solar corona and at interplanetary shocks, have 
recently been analyzed in terms of superdiffusive transport models
\citep{Perri-Zimbardo-2007, Perri-Zimbardo-2009, Sugiyama-Shiota-2011,
  Trotta-Zimbardo-2011, Zimbardo-Perri-2013}.  Notably, those studies
relied on an asymptotic expression for a non-Gaussian propagator
(equivalent to the Green's function of a fractional diffusion
equation), which has a limited validity range.

In this paper we develop more accurate analytical solutions and argue 
that they should be used to put the superdiffusive particle transport 
in the heliosphere on a firmer footing. As a concrete example, 
we examine a prototypical transport problem, described by 
a one-dimensional fractional diffusion-advection equation (Section 2). 
We derive an exact solution by the Fourier transform (Section 3), 
and we obtain an approximate solution in terms of elementary functions 
in a weak diffusion approximation (Section 4). 
We demonstrate the accuracy of the approximation by comparing the new solution 
with a semi-numerical solution based on a formally exact Fourier series 
expansion (Section 5), and we discuss what our results imply for 
the interpretation of the observed particle distributions as evidence 
of superdiffusive transport in the heliosphere (Section 6).

\section{Formulation of the problem}
Consider first the usual diffusion-advection equation 
for a one-dimensional particle distribution function $f(x,t)$ 
that depends on position $x$ and time $t>0$: 
\begin{equation}
  \frac{\partial f}{\partial t} = 
  \kappa \frac{\partial^2 f}{\partial x^2} +
  a \frac{\partial f}{\partial x} + \delta(x) . 
\end{equation}
Here $a$ is a constant advection speed, and $\kappa$ is a constant
diffusion coefficient.  This transport equation follows from the Fokker--Planck 
equation for energetic particles if their energy losses are
neglected and pitch-angle scattering is strong \citep[e.g.,][and
references therein]{Schlickeiser-Shalchi-2008,
  Litvinenko-Schlickeiser-2013}. A delta-functional source term on the
right may correspond to energetic particles injected at a shock, and
$a$ can be interpreted as the background solar wind speed.

Standard methods \citep[e.g.,][]{Carslaw-Jaeger-1959} give the
solution of the initial value problem $f(x, 0) = 0$ 
on the interval $-\infty < x < \infty$:
\begin{equation}
  f(x, t) = 
  \int_0^t (4 \pi \kappa t')^{-1/2} 
  \exp \left[ -\frac{(x+at')^2}{4 \kappa t'} \right] dt' . 
\end{equation}
A steady distribution is established in the limit $t \to \infty$: 
\begin{equation}
  f(x, \infty) = \frac{1}{a} \exp \left( -\frac{a}{\kappa} x \right) , 
  \quad x>0 , 
  \label{eq-gauss-steady}
\end{equation} 
\begin{equation}
  f(x, \infty) = a^{-1} , \quad x<0 . 
\end{equation}

In the remainder of the paper, we investigate the following fractional
differential equation that generalizes the usual diffusion-advection
equation:
\begin{equation}
  \frac{\partial f}{\partial t} = 
  \kappa \frac{\partial^{\alpha} f}{\partial |x|^{\alpha}} + 
  a \frac{\partial f}{\partial x} + \delta(x) 
\label{eq-frac-adv-diff}
\end{equation}
\citep[e.g.,][]{Stern-etal-2014}. The equation governs the evolution 
of a distribution function $f(x, t)$ for $t>0$ and $-\infty < x < \infty$. 
For simplicity, we assume the initial condition 
\begin{equation}
  f(x, 0) = 0 . 
\end{equation} 
The advection speed $a$ and diffusion coefficient $\kappa$ 
(now with dimensions length$^\alpha$/time) are positive constants. 
We use the Riesz derivative to define a fractional spatial derivative: 
\begin{align}
  \frac{\partial^{\alpha} f(x,t)}{\partial |x|^{\alpha}} &= 
  \frac{1}{\pi} 
  \sin \left( \frac{\pi}{2} \alpha \right) 
  \Gamma (1+\alpha) \nonumber\\
  &\times\int_0^{\infty} \frac{f(x+\xi)-2f(x)+f(x-\xi)}{\xi^{1+\alpha}} d\xi
\end{align}
\citep[][]{Samko-etal-1993,Saichev-Zaslavsky-1997}. 
Because the derivative corresponds to a fractional Laplacian
operator in higher dimensions, an alternative notation $-(-\Delta)^{\alpha/2} f$ 
is also used \citep{Mainardi-etal-2001}. Although the regularized form
above is defined for $0 < \alpha < 2$, in what follows we are mainly
interested in the superdiffusive case $1 < \alpha < 2$.

\section{Fourier transform solution and asymptotics}
The Fourier transform gives a convenient method of 
solving the fractional diffusion-advection equation 
on the interval $-\infty < x < \infty$. 
Taking the Fourier transform of equation (\ref{eq-frac-adv-diff}) yields 
\begin{equation}
  \frac{\partial \tilde{f}}{\partial t} = 
  -\kappa |k|^{\alpha} \tilde{f}  
  + iak \tilde{f}  + \frac{1}{2 \pi} , 
\end{equation}
where 
\begin{equation}
  \tilde{f} (k, t) = \frac{1}{2 \pi} 
  \int_{-\infty}^{\infty} f (x, t) \exp(-ikx) dx . 
\end{equation}
Integration of the first-order equation for $\tilde{f}$ yields  
\begin{equation}
  \tilde{f} (k, t) = 
  \frac{1}{2 \pi} 
  \frac{1 - \exp [(iak - \kappa |k|^{\alpha}) t]}{\kappa |k|^{\alpha} - iak} , 
\end{equation}
where an integration constant is specified by the initial condition 
$\tilde{f} (k, 0) = 0$.
Now the inverse Fourier transform gives the solution 
\begin{equation}
  f(x, t) = 
  \frac{1}{2 \pi} \int_{-\infty}^{\infty} 
  \frac{1 - \exp [(iak - \kappa |k|^{\alpha}) t]}{\kappa |k|^{\alpha} - iak} 
  \exp(ikx) dk . 
\label{eq-f-fourier}
\end{equation}
In the non-diffusive case, $\kappa = 0$ and the integral is evaluated
to give an expanding ``top-hat'' solution
\begin{equation}
  \left. f (x, t) \right|_{\kappa=0} = f_0 (x, t) = 
  \frac{1}{2a} \left[ \mbox{sgn}(x+at) - \mbox{sgn}(x) \right] . 
\label{eq-f-kappa0}
\end{equation}

The solution of equation (\ref{eq-frac-adv-diff}) can also be expressed as 
\begin{equation}
  f(x, t) = \int_0^t G(x+at', t') dt' , 
\label{eq-f-int-G}
\end{equation}
where the Green's function $G (x, t)$ 
satisfies the fractional diffusion equation 
\begin{equation}
  \frac{\partial G}{\partial t} = 
  \kappa \frac{\partial^{\alpha} G}{\partial |x|^{\alpha}} + 
  \delta(x) \delta(t) . 
\end{equation}
Its Fourier transform is given by 
\begin{equation}
  \frac{\partial \tilde{G}}{\partial t} = 
  -\kappa |k|^{\alpha} \tilde{G}  
  + \frac{1}{2 \pi} \delta(t) . 
\end{equation}
It follows that 
\begin{equation}
  \tilde{G} (k, t) = \frac{1}{2 \pi} 
  \exp(-\kappa |k|^{\alpha} t) , 
\end{equation}
and the inverse Fourier transform yields  
\begin{equation}
  G(x, t) = 
  \frac{1}{\pi} \int_{0}^{\infty} 
  \exp(-\kappa k^{\alpha} t) \cos (kx) dk 
\label{eq-Green}
\end{equation}
\citep[e.g.,][]{Chukbar-1995}. 

An asymptotic expression for $G(x, t)$ is obtained by integrating 
equation (\ref{eq-Green}) by parts and applying an analogue of Watson's lemma 
for Fourier type integrals \citep[e.g.,][]{Ablowitz-Fokas-1997}. 
For $x \gg (\kappa t)^{1/\alpha}$, the result is 
\begin{equation}
  G(x, t) \approx 
  \frac{1}{\pi} 
  \sin \left( \frac{\pi}{2} \alpha \right) 
  \Gamma (1+\alpha) 
  \frac{\kappa t}{|x|^{1+\alpha}} . 
\label{eq-G-small-t}
\end{equation}
For $x > 0$, substitution into equation (\ref{eq-f-int-G}) yields 
\begin{equation}
  f(x, t) \approx 
  \frac{1}{\pi} 
  \sin \left( \frac{\pi}{2} \alpha \right) 
  \Gamma (1+\alpha) 
  \int_0^t \frac{\kappa t'}{(x+at')^{1+\alpha}} dt' , 
\end{equation}
and so the distribution function $f(x, t)$ is given by 
\begin{equation}
  f(x, t) \approx 
  \frac{1}{\pi} 
  \sin \left( \frac{\pi}{2} \alpha \right) 
  \Gamma (\alpha-1) 
  \frac{\kappa}{a^2}
  \left[ x^{1-\alpha} - 
  \frac{x+\alpha a t}{(x+at)^{\alpha}} \right] 
\label{eq-f-x>0}
\end{equation}
as long as $ x+at \gg (\kappa t)^{1/\alpha} $ and $1 < \alpha < 2$. 
Two limiting cases are as follows:
\begin{align}
  \label{eq-f-small-t}
  f(x, t) &\approx 
  \frac{1}{2 \pi} 
  \sin \left( \frac{\pi}{2} \alpha \right) 
  \Gamma (1+\alpha) 
  \frac{\kappa t^2}{x^{1+\alpha}} , 
  \quad x \gg at , \\
  \label{eq-f-large-t}
  f(x, t) &\approx 
  \frac{1}{\pi} 
  \sin \left( \frac{\pi}{2} \alpha \right) 
  \Gamma (\alpha-1) 
  \frac{\kappa}{a^2} x^{1-\alpha} , 
  \quad 0 < x \ll at . 
\end{align}

Equation (\ref{eq-f-large-t}) is essentially the asymptotic power-law 
that was previously used in the data analysis of energetic particles,
accelerated in the solar corona \citep{Trotta-Zimbardo-2011} and at
interplanetary shocks \citep{Perri-Zimbardo-2007, Perri-Zimbardo-2009,
  Sugiyama-Shiota-2011}. For instance, \citet{Perri-Zimbardo-2007,
  Perri-Zimbardo-2009} used a formula for the particle density at a
fixed point due to a shock-associated source moving with a speed
$V_{\rm sh}$. The formula follows from our analysis by changing the
reference frame. Suppose a shock, initially located at
$x_0 = - V_{\rm sh} t_0$, moves at speed $V_{\rm sh}$ and arrives at 
$x=0$ when $t=0$.  On changing variables, equation (\ref{eq-f-x>0}) yields
\begin{align}
\label{eq-Perri-2007}
  f(x, t) &\approx 
  \frac{1}{\pi} 
  \sin \left( \frac{\pi}{2} \alpha \right) 
  \Gamma (\alpha-1) 
  \frac{\kappa}{V_{\rm sh}^2} \\
  &\times\left[ 
  (x - V_{\rm sh} t)^{1-\alpha} - 
  \frac{x + (\alpha - 1) V_{\rm sh} t + \alpha V_{\rm sh} t_0}{(x+ V_{\rm sh} t_0)^{\alpha}} 
  \right] \nonumber\,,
\end{align} 
which reduces to equation (4) in \citet{Perri-Zimbardo-2007} on
assuming that the shock is coming from a very large distance, $t_0 \to
\infty$. Note for clarity that \citet{Perri-Zimbardo-2007} do not
specify a normalization constant $b$ in their equation (1), and
their notation is different: their $\mu$ is our $1 + \alpha$, and
their $\alpha$ is our $3 - \alpha$.  The latter expression appears in
the dependence of the variance of a particle displacement on time
$\sim t^{3 - \alpha}$ when a finite particle speed is taken into
account.

\section{A weak diffusion approximation}

A limitation of the analysis in the previous section is that we used a
short-time asymptotic (\ref{eq-G-small-t}) for the Green's function
$G(x,t)$ to derive equation (\ref{eq-f-x>0}) for the particle
distribution function $f(x, t)$, and so it is not clear whether the
results are valid for $at \gg x$.  In addition, the
results are only valid for $x>0$ because the integral in equation 
(\ref{eq-f-int-G}) diverges for $x<0$ if equation (\ref{eq-G-small-t})
is used to evaluate the integral.  To remove these limitations of the
analysis, we solve for $f(x,t)$ in a weak diffusion approximation 
that allows us to evaluate the integral 
in equation (\ref{eq-f-fourier}) for both $x>0$ and $x<0$.

The superdiffusive term in equation (\ref{eq-frac-adv-diff}) 
can be treated as a perturbation sufficiently far from the locations 
where the advective solution (\ref{eq-f-kappa0}) predicts jumps in $f(x,t)$. 
Suppose $l_d$ is the distance from a jump where 
the diffusive and advective terms in equation (\ref{eq-frac-adv-diff}) 
are comparable. To an order of magnitude, 
$\kappa \partial^{\alpha} f / \partial |x|^{\alpha} \sim \kappa f / l_d^{\alpha}$ and 
$a \partial f / \partial x \sim a f / l_d$. The diffusion length is thus defined as 
\begin{equation} 
  l_d = ( \kappa / a )^{1/(\alpha-1)} . 
\end{equation} 
Now assuming that 
\begin{equation} 
  l_d \ll |x|, |x+at|, at , 
\end{equation}
we can formally treat $\kappa$ as a small parameter, and so 
\begin{equation}
  f(x, t) \approx f_0 (x, t) + 
  \kappa \left. \frac{df(x, t)}{d\kappa} \right|_{\kappa=0} , 
\end{equation}
where $f_0 (x, t)$ is given by equation (\ref{eq-f-kappa0}). 
Differentiation of equation (\ref{eq-f-fourier}) with respect to $\kappa$ yields 
\begin{align}
  f(x, t) &\approx f_0 (x, t) \nonumber\\
  &+\frac{\kappa}{\pi a^2} \int_{0}^{\infty} 
    \biggl[  \frac{\cos kx - \cos k(x+at)}{k^{2-\alpha}} \nonumber\\
  &- \frac{(\alpha-1)at}{(x+at)} \frac{\cos k(x+at)}{k^{2-\alpha}} \biggr] dk , 
\end{align}
where the last term in the integrand is obtained by integrating by parts 
and neglecting a rapidly varying term containing $\exp[ik(x+at)]$ 
at the upper integration limit. On simplifying and using Watson's lemma, 
we get 
\begin{align}
  f(x, t) &\approx f_0 (x, t) + \frac{1}{\pi} 
  \sin \left( \frac{\pi}{2} \alpha \right) 
  \Gamma (\alpha-1) 
  \frac{\kappa}{a^2}\nonumber\\
  &\times\left[ |x|^{1-\alpha} - 
  \frac{x+\alpha a t}{(x+at) |x+at|^{\alpha-1}} \right] .
\label{eq-weak-diff}
\end{align}

For $x>0$, we recover equation (\ref{eq-f-x>0}) and its limiting cases, confirming 
the analysis of the previous section. For $x<0$, we have 
\begin{equation}
  f(x, t) \approx 
  \frac{1}{2 \pi} 
  \sin \left( \frac{\pi}{2} \alpha \right) 
  \Gamma (1+\alpha) 
  \frac{\kappa t^2}{|x|^{1+\alpha}} , 
  \quad |x| \gg at , 
\end{equation}
\begin{equation}
  f(x, t) \approx \frac{1}{a} + 
  \frac{1}{\pi} 
  \sin \left( \frac{\pi}{2} \alpha \right) 
  \Gamma (\alpha-1) 
  \frac{\kappa}{a^2} |x|^{1-\alpha} , 
  \quad |x| \ll at . 
  \label{eq-full-asymptotic}
\end{equation}
The discontinuity at $x=-at$ broadens into a smoother transition. 
The solution is inapplicable in the vicinity of the jump at $|x+at|\approx 0$, 
however, because the weak diffusion approximation is valid only as long as 
$|x|$, $|x+at|$, and $at$ are large in comparison with the diffusion length $l_d$, 

\section{Accuracy of the approximation}
\label{sec:series}
\citet{Stern-etal-2014} gave the following exact Fourier series solution 
to equation (\ref{eq-frac-adv-diff}) on a domain of length $L$: 
\begin{align}
\label{eq-fourier-series}
 f(x,t) &= 
\sum_{n=1}^{\infty}\Biggl\{(1+(-1)^{n+1}) \biggl[ \left(\frac{n \pi}{L}\right)^{\alpha} \kappa L \cos\left(\frac{n \pi x}{L} \right)\nonumber\\
&-n \pi a \sin\left(\frac{n \pi}{L}x \right) \nonumber \\
&-\left(\frac{n \pi}{L} \right)^{\alpha} \kappa L \cos\left(\frac{n \pi}{L}(x+at) \right) \exp\left(-\left(\frac{n \pi}{L} \right)^{\alpha} \kappa t \right) \nonumber \\
& + n \pi a \sin\left( \frac{n \pi}{L} (x+at) \right) \exp\left(-\left(\frac{n \pi}{L} \right)^{\alpha} \kappa t \right) \biggr] \text{\Huge{/}} \nonumber \\
& \left(  \left(\frac{n \pi}{L}\right)^{2 \alpha}\kappa^2L^2 + n^2 \pi^2a^2  \right) \Biggr\} \,.
\end{align}
Because $f(x,t) = f(x+2L, t)$, the series does not represent 
the solution of an initial-value problem on an infinite interval 
for $t\rightarrow\infty$. For a localized source and finite $t$, however, 
the series solution accurately represents $f(x,t)$ 
on an infinite interval if $L$ is sufficiently large. In practice, 
we achieve accuracy by choosing $L \gg at$. 

We compare the new analytical solutions of the previous sections with
a semi-numerical solution based on the Fourier series expansion. We
use equation~(\ref{eq-fourier-series}) and sum up $N=10^6$ terms with
$L=1000$ to achieve high accuracy. We set $a=1$, which simply means
that in what follows we measure speeds in units of the solar wind
speed. We choose $\alpha=1.5$ in agreement with the range of values
inferred from the heliospheric particle data
\citep{Perri-Zimbardo-2007, Perri-Zimbardo-2009,
  Sugiyama-Shiota-2011}.  The superdiffusion coefficient $\kappa$ is a
key parameter of the theory. Work is underway to estimate $\kappa$
from the data (Perri et al., in preparation).  We adopt $\kappa=0.5$
as an illustration and investigate how the particle distribution
evolves over a few hundred advection times.

\begin{figure}
\noindent\includegraphics[width=0.48\textwidth]{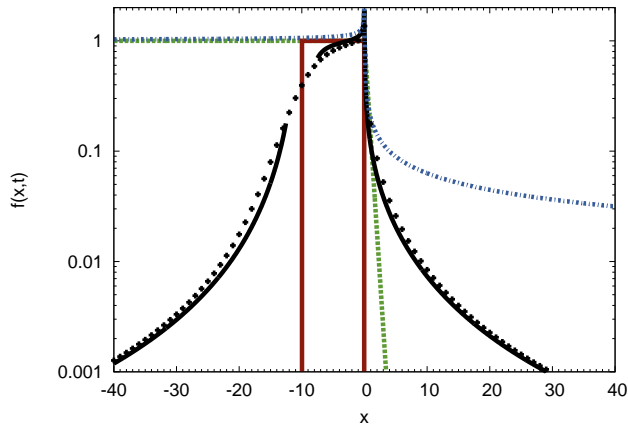}
\caption{The Fourier transform solution in a weak diffusion limit
  (equation~(\ref{eq-weak-diff}), solid black line) and the series
  solution (equation~(\ref{eq-fourier-series}), black symbols) at time
  $t=10$. The dot-dashed blue line gives the approximate steady solution 
  in equation~(\ref{eq-f-large-t}) for $x>0$ and equation~(\ref{eq-full-asymptotic}) 
  for $x<0$. For reference, the dashed green line shows the steady state Gaussian
  diffusion solution in equation~(\ref{eq-gauss-steady}), and the red box shows 
  the expanding top-hat non-diffusive solution in equation~(\ref{eq-f-kappa0}). 
  Parameters are $\alpha=1.5$, $\kappa=0.5$, $a=1$.}
\label{fig:fourier-t10}
\end{figure}

Figure~\ref{fig:fourier-t10} shows the results 
in a semi-logarithmic plot at time $t=10$. We find a good
agreement between the analytical (solid black line) and semi-numerical
(black symbols) solutions, except for the region around $x=-at$ where
the weak diffusion approximation is not valid. The red box in the
figure illustrates the non-diffusive solution given by 
equation~(\ref{eq-f-kappa0}). We have truncated the analytical solution
in the vicinity of $x=-at$ over a length $l$
of ten times the diffusion length ($l=10\,l_d=2.5$). The green and blue lines
give the steady-state solution for the Gaussian diffusion case, 
given by equation~(\ref{eq-gauss-steady}), and the approximate steady-state
solution, given by equation~(\ref{eq-full-asymptotic}).

\begin{figure}
\noindent\includegraphics[width=0.48\textwidth]{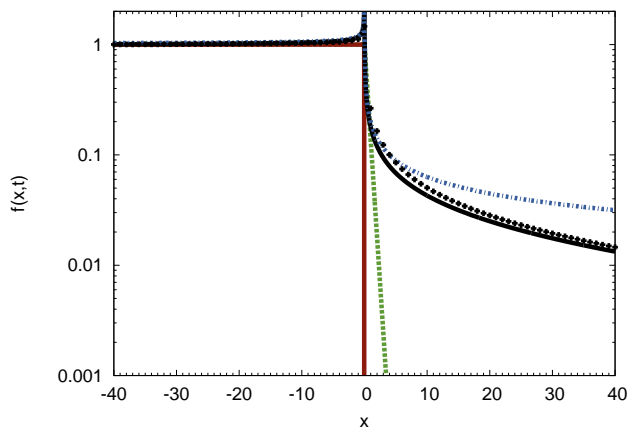}
\caption{Same as Figure~\ref{fig:fourier-t10} but now at time
  $t=200$.}
\label{fig:fourier-t200}
\end{figure}
Figure~\ref{fig:fourier-t200} gives the same solutions as in
Figure~\ref{fig:fourier-t10} at a later time $t=200$. 
The weak diffusion solution and the Fourier series remain in good agreement 
as they slowly approach the steady state. The downstream region 
in Figure~\ref{fig:fourier-t200} is already completely filled since $|x| \ll at$. 
An interesting feature of the solution is a peak at the injection site $x=0$, 
which is not present for Gaussian diffusion.

\section{Discussion}
We used the Fourier transform to solve analytically a fractional
diffusion-advection equation for cosmic ray transport, and we applied the solution 
to the problem of describing the transport of energetic particles, 
accelerated at a traveling heliospheric shock. 
We also developed a weak diffusion approximation, 
based on the exact Fourier transform solution. 
We confirmed the validity of the approximation for both
early and late times by comparing it with an exact Fourier series solution. 
Our analysis is motivated by 
recent applications of superdiffusive transport models to 
the observed shock-accelerated particle distributions 
\citep{Perri-Zimbardo-2007, Perri-Zimbardo-2009, Sugiyama-Shiota-2011}.

Our new solution quantifies the limited validity of the
asymptotic expressions, used previously to interpret the particle
data. Specifically, the formula used by \citet{Perri-Zimbardo-2007,
  Perri-Zimbardo-2009} and \citet{Sugiyama-Shiota-2011} is basically
our equation~(\ref{eq-Perri-2007}) in the limit $t_0 \to \infty$,
corresponding to a shock approaching an observer from a very large
distance $V_{\rm sh} t_0$.  As our results show, however, it may take
a very long time for the asymptotic expression to become accurate. 
The ratio of the second term in equation (\ref{eq-Perri-2007}) 
to the first one is $(-t/t_0)^{\alpha-1} (\alpha + (\alpha-1) t/t_0)$ 
at $x=0$, and so our more accurate solution differs from the $t_0 = \infty$ 
asymptotic expression by about a factor of 2 when the distance between 
the observer and the shock is as short as one tenth 
of the initial distance $V_{\rm sh} t_0$ between them.

To sum up, solar cosmic-ray data in various settings appear to be
consistent with asymptotic propagator solutions to a fractional
diffusion equation or more general continuous-time random-walk models
\citep{Zimbardo-Perri-2013}. We argued, however, that more accurate solutions 
of an appropriate transport equation should be used for validating the
superdiffusive transport of energetic particles in the heliosphere. 
In the context of the transport of particles, accelerated at 
a traveling heliospheric shock, our analysis strongly suggests that 
we should not assume the initial distance $V_{\rm sh} t_0$ 
of the shock from the observer to be infinite. The shock travel time 
$t_0$ should be a parameter of the superdiffusive transport model.

\acknowledgments{Y.L. thanks Horst Fichtner for drawing his
  attention to the topic of this paper. F.E. appreciates support from
  the International Space Science Institute (ISSI) during a team
  meeting in Bern on superdiffusive transport and thanks the team members 
  for many insightful discussions. The authors acknowledge 
  an anonymous referee for several useful suggestions.}


\end{document}